# Title: Molecular-Scale Visualization of Steric Effects of Ligand Binding to Reconstructed Au(111) Surfaces

**Authors:** Liya Bi[1,2], Sasawat Jamnuch[3], Amanda Chen[3], Alexandria Do[2,3], Krista P. Balto[1], Zhe Wang[4], Qingyi Zhu[1], Yufei Wang[2,3], Yanning Zhang[4], Andrea R. Tao[1,2,3]\*, Tod A. Pascal[2,3]\*, Joshua S. Figueroa[1,2]\*, Shaowei Li[1,2]\*

**Affiliations:**
[1]Department of Chemistry and Biochemistry, University of California, San Diego; La Jolla, CA 92093-0309, USA.
[2]Program in Materials Science and Engineering, University of California, San Diego; La Jolla, CA 92093-0418, USA.
[3]Department of Nano and Chemical Engineering, University of California, San Diego; La Jolla, CA 92093-0448, USA.
[4]Institute of Fundamental and Frontier Sciences, University of Electronic Science and Technology of China; Chengdu, 611731, China.
\*Corresponding author. Email: atao@ucsd.edu (A.R.T.); tpascal@ucsd.edu (T.A.P.); jsfig@ucsd.edu (J.S.F.); shaoweili@ucsd.edu (S.L.)

**Abstract:** Direct imaging of single molecules at nanostructured interfaces is a grand challenge, with potential to enable new, precise material architectures and technologies. Of particular interest are the structural morphology and spectroscopic signatures of the adsorbed molecule, where modern probes are only now being developed with the necessary spatial and energetic resolution to provide detailed information at molecule-surface interface. Here, we directly visualize the binding of individual m-terphenyl isocyanide ligands to a reconstructed Au(111) surface through scanning tunneling microscopy (STM) and inelastic electron tunneling spectroscopy (IETS). The site-dependent steric pressure of the various surface features alters the vibrational fingerprints of the m-terphenyl isocyanides, which is characterized with single-molecule precision through joint experimental and theoretical approaches. This study for the first time provides molecular-level insights into the steric-pressure-enabled surface binding selectivity, as well as its effect on the chemical properties of individual surface-binding ligands.

**One-Sentence Summary:** The microscopic visualization at a single-molecule level illustrates how steric effects alter the properties of m-terphenyl isocyanides when binding to Au(111) surfaces.

**Main Text:**
Steric interference refers to the repulsion between atoms or functional groups that are forced together by geometric limitations within or between molecules.(*1-4*) Its origin is in the fundamental Pauli force, as well as long-range interactions between the dipole moments of adjacent atoms or functional groups,(*5-8*) which creates a destabilizing force that can distort a molecule from its preferred geometry,(*9, 10*) induce dissociation or bond-breaking events,(*11*) or inhibit chemical reactivity.(*12-14*) Accordingly, the concept of steric interference has long-standing importance in the design, study and understanding of chemical(*1-4*) and biological(*15*) systems. Indeed, the exploitation of steric interference has been central to the isolation of inherently reactive molecules,(*4, 16-18*) and has underpinned the design of catalysts for asymmetric synthesis.(*19*) Maximizing or minimizing steric interferences are also critical design principles for selective drug delivery to enzyme active sites.(*20*) In addition, steric interference — whether in an intramolecular or intermolecular context — can significantly affect the conformational preferences of a molecule(*21*) or the interactions between molecules,(*22*) which in turn can impact physical properties such as melting point, boiling point, and solubility.

From a synthetic standpoint, modulation of steric properties can provide an accessible control knob to achieve chemical selectivity. A molecule with designed steric features can preferably interact with other substrates in areas where steric repulsion is minimized.(*23*) The ability to control reactions sites or regions can also enable the selective and predictable adsorption of molecules onto structured surfaces. This concept can be particularly useful in the field of nanotechnology, where the properties of nanostructures can be tuned by modifying their surface chemistry through the use of ligands.(*24-29*) For example, the selective adsorption of ligands onto different facets of a nanocrystal can control its shape and size and, as a result, directly influence its optical, electronic, and catalytic properties.(*30-33*) Furthermore, the potential to harness steric effects in the design of ligands that specifically target certain surface sites on a nanostructure would have a profound impact on the creation of highly selective and efficient catalysts for a range of chemical reactions.(*34-37*) Therefore, utilizing, and ultimately controlling, steric effects is crucial for manipulating properties of nanostructures and developing new materials with tailored functionalities.

When designing a system that is sensitive to ligand-surface steric interactions, it is important that the surface-binding group possesses sites of potential steric interference that are well defined. *m*-Terphenyl isocyanides (**Fig. 1A**) are a promising class of ligands in this regard due to their strong surface binding ability and their unique and modifiable steric profile.(*38-41*) These ligands comprise an aryl isocyanide (i.e. CNAr; Ar = aryl) binding group, which has been long established to bind to metal surfaces.(*42-47*) In the m-terphenyl modification, the isocyanide group is flanked by two additional, mutually-meta, encumbering arenes, which create significant steric interference and pressure in the direction pointing towards the metal surface. This steric pressure is maximized when the m-terphenyl isocyanide ligand is bound to a planar metal surface. However, it can be significantly reduced when the ligand is bound to a convex surface, such as the step edge on a metal surface, where the m-terphenyl group can localize in a less sterically hindered binding environment. Recent spectroscopic studies have provided evidence of preferable adsorption of m-terphenyl isocyanide ligands to nanocrystal surfaces exhibiting high degrees of nanocurvature.(*23*) Specifically, it was shown that the m-terphenyl isocyanide (**Fig. 1A**), $CNAr^{Mes2}$ ($Ar^{Mes2}$ = 2,6-(2,4,6-$Me_3C_6H_2$)$_2C_6H_3$), could readily bind to Au nanospheres (AuNS) with diameters between 5–50 nm, but did not bind to larger diameter particles with lower-degrees of nanocurvature to an appreciable extent. This sterically-induced binding selectivity enabled the development of a chemical method for nanoparticle separation based on size as well as a chemical means of effecting nanoparticle size-focusing. However, while this ensemble-level study elucidated the global effects of $CNAr^{Mes2}$ binding to AuNS surfaces, direct visualization of this binding, as well as information concerning the precise steric interactions between the ligands and their nanoscale environment were absent. Here, we fill in this gap with molecular-scale characterization of steric-pressure-induced site-selective binding of individual m-terphenyl isocyanide ligands to a reconstructed Au(111) surface using scanning tunneling microscopy (STM),(*48, 49*) inelastic electron tunneling spectroscopy (IETS)(*50, 51*) and computational simulations. The results presented here provide a detailed structural and spectroscopic picture of the role of steric effects at the ligand-surface interface.

**Visualization of Selective Binding of $CNAr^{Mes2}$ to Au(111)**

Here, we study the binding of $CNAr^{Mes2}$ to the reconstructed Au(111) surfaces. We choose Au(111) as the substrate because of its diverse surface structures, including high-curvature step edges as well as alternating face center cubic (FCC) and hexagonal close-packed (HCP) facets separated by protruding herringbone reconstructions.(*52, 53*) As shown in **Fig. 1, B and C**, each topological site possesses inherently distinct nano curvature that can be resolved by STM topographic imaging. In addition, the distinct curvature of these sites leads to an inhomogeneous steric environment upon molecular adsorption. Importantly, both the step edges and herringbones sites have convex curvature, which is expected to exhibit low degrees of steric pressure on bound ligands.

To understand the baseline profile of $CNAr^{Mes2}$ adsorption, atomically clean Au(111) surfaces were dosed with evaporated ligand at 5 K and $10^{-10}$ Torr in situ at the STM junction. Initial STM topographical images taken at this temperature (**Fig. 1B**) revealed that the $CNAr^{Mes2}$ ligands are adsorbed randomly on the Au(111) surface without any discernible preference for a particular site. Even at very low molecular coverage, isolated $CNAr^{Mes2}$ molecules can be seen on both the FCC and HCP basal plane between herringbones, protruding at a height of about 250 pm (**Fig. 1D**). At such a low temperature, it is expected that the $CNAr^{Mes2}$ molecules rapidly release thermal energy to the environment and remain close to the location of initial deposition. To probe for thermally-induced migration events — especially to energetically favorable adsorption sites — the Au(111) substrate with sub-monolayer $CNAr^{Mes2}$ coverage was brought to room temperature and subsequently re-cooled to 5 K . Topographical images taken afterwards indicated clear migration of the $CNAr^{Mes2}$ molecules within the totality of the substrate (**Fig. 1E**). While some $CNAr^{Mes2}$ ligands can be found on the herringbone elbow sites after annealing, the vast majority of the molecules migrate to the step-edge positions in both the FCC and HCP domains. Notably, previous research has identified herringbone elbows as the most chemically reactive sites for ligand binding on Au(111) surfaces, followed by FCC domains.(*54-56*) In contrast, HCP facets and step edges both exhibit relative inertness. However, our findings clearly indicate a preference of CNAr$^{Mes2}$ ligands for adsorbing on HCP step edges over herringbone elbows (as shown in the inset of **Fig. 1E**). This observation strongly suggests the presence of an energetic factor beyond metal-ligand binding considerations, which is attributed to reduced steric repulsion resulting from the large convex surface curvature at the edge sites and supported by theoretical calculations. As shown in **Fig. 1F**, the adsorption energy of the molecule on the basal plane is nearly negligible, while a ~ 1.2 eV deep potential well exists at the Au edge **(Fig. 1G)**, and is characterized by a ~ 2 Å long Au-C bond. This calculated Au-C bond length is consistent with those found for structurally characterized molecular gold-isocyanide complexes (average d(Au-C) = 1.964 +/- 0.037Å; **fig. S3B**).(*57*) Moreover, we also observed molecular clusters on the FCC domains after annealing, indicating that the intermolecular interaction could also alter either the steric pressure or the ligand-metal interaction which in turn affect the molecular binding behavior.

**Influence of Steric Effects on Adsorption Structure of CNAr$^{Mes2}$ on Au(111)**
The STM images of individual CNAr$^{Mes2}$ ligands adsorbed at different sites provide additional evidence of the influence of steric effects on the molecular adsorption structure and the migration kinetics. Shown in **Fig. 2, A-C** are the topographic images of the isolated CNAr$^{Mes2}$ ligands on the herringbone elbow site, at the step edge, and on the FCC basal plane respectively (see also isolated CNAr$^{Mes2}$ on the HCP basal plane in **fig. S4**). In **Fig. 2A**, the molecule is situated on top of the herringbone elbow site and appears crescent-shaped, less symmetric than its molecular structure (**Fig. 1A**). This indicates that the molecule tilts to one side upon its adsorption on the herringbone elbow site, in good agreement with the ~10º tilt angle given by computational simulations (**fig. S5A**). At the step edge, the molecule appears to straddle the edge with one mesityl group on the upper Au layer and the other on the lower layer (**Fig. 2B**), with a calculated tilting angle of ~ 32º relative to the (111) direction (**fig. S5B**). We find that molecular adsorption at the step edge is highly stable, while adsorption on top of the herringbone elbow site can change conformations when disturbed by the STM tip (**fig. S6**). This observation agrees with the notion that the herringbone elbow sites, which possess an intermediate degree of surface curvature, also present greater ligand-surface steric pressures than the step edge sites. Such a steric-bulk-induced bond weakening is most evident in the absence of any convex curvature. The images of isolated CNAr$^{Mes2}$ ligands adsorbed on both FCC and HCP basal planes show a six-lobe feature (**Fig. 2C and fig. S4**), presenting a rapidly switching/rotating behavior among six equivalent adsorption geometries which are defined by the symmetry of Au(111). This is consistent with our molecular dynamics (MD) simulations at 5K, which shows a freely rotating CNAr$^{Mes2}$ on the planar surface, but only vibrating molecules on the herringbones and at the step edges (**movies S1 – S3**). Importantly, these simulations reveal that the steric pressure between the CNAr$^{Mes2}$ ligands and the surface perturbs the binding to such an extent that, even at a low temperature of 5 K, thermal energy is sufficient to excite rotational motion around the metal-binding isocyanide group. This was further confirmed by calculating the in-plane and out-of-plane rotational temperatures from MD simulations (**table S1**), where we find populated low energy rotational states for the molecule on the planar surface at 5K, which are not populated for molecules at the herringbone or step edges.

**Imaging Surface-Induced Vibrational Modes of CNAr$^{Mes2}$ on Au(111)**

Remarkably, the spectroscopic properties of the individual CNAr$^{Mes2}$ ligands are sensitive to the variation in steric pressures at its unique surface binding environment and is evident from modifications on their molecular vibrational fingerprints. To probe the vibrational features of CNAr$^{Mes2}$ ligands at different binding sites, inelastic electron tunneling spectroscopy (IETS)(*50, 51*) was utilized, which is a highly efficient technique for investigating low-energy molecular vibrations at a sub-molecular scale. The low-energy vibrational modes are highly sensitive to the molecule-surface interaction, and thus are a valuable means of characterizing the variations in molecular properties that occur in response to local chemical environments. IETS measures the second derivative of the tunneling current with respect to bias ($d^2I/dV^2$). A pair of symmetric peak and dip observed over the origin of the spectra indicates the bias corresponding to an inelastic excitation, such as molecular vibration,(*50, 58*) rotation,(*59, 60*) or spin excitation.(*61, 62*) The $d^2I/dV^2$ spectra obtained from an isolated CNAr$^{Mes2}$ ligand on the FCC basal plane exhibits rich vibrational features below 85 meV/685.6 cm$^{-1}$ with the intensity of the peaks and dips varying within the molecule (labeled with I-VI in **Fig. 3A**). Since STM-IETS detects the conductance change of the junction due to the vibrational excitation, its spatial distribution closely resembles the nuclear motions. **Fig. 3A** depicts that the signal emanating from the ~49 meV and ~67 meV vibrational modes (modes V and VI) are highly conspicuous and away from the center of the molecule (green spectrum). Conversely, the lower energy modes between ~18 meV and ~34 meV (modes I-IV) exhibit a more robust signal intensity near the center of the molecule (black spectrum). The spatially resolved mappings of the $d^2I/dV^2$ signal (**Fig. 3, B-G and fig. S7**) provide an intuitive microscopic visualization of the vibrational motions. The images captured at 23.9 mV, 28.6 mV, and 33.4 mV (modes II, III and IV) exhibit a bullseye feature, which provides a distinctive motion pattern involving both the central aryl ring and the two para-methyl groups. In contrast, the images taken at 19.1 mV, 48.5 mV, and 67.7 mV (modes I, V, and VI) display a donut shape whose diameter, ~ 7 Å, is close to the separation between carbon atoms in the ortho-methyl groups on different mesityl groups of a free-standing CNAr$^{Mes2}$, indicating the primary motion of four ortho-methyl groups. It is worth mentioning that the $d^2I/dV^2$ images taken with biases below 30 mV do not closely follow the molecular symmetry. This phenomenon can be attributed to the coupling of vibrational excitation with the surface state of Au(111). Specifically, the Friedel oscillation resulting from the scattering of low-energy electrons near Au Fermi level leads to a spatial variation in the electron density of states,(*63, 64*) which in turn breaks the symmetry of the excitation cross section of the molecular vibration.

  By comparing the microscopic patterns with theoretical simulation (**Fig. 4, A-H and fig. S8**), we can clearly identify the detailed molecular motion corresponding to the experimentally observed vibrational modes. Due to the structure complexity of the CNAr$^{Mes2}$ ligand, rich features are exhibited in the simulated vibrational density (**Fig. 4A**). Six vibrational modes stand out in the simulated tunneling current (**Fig. 4B**) due to their relatively large out-of-plane nuclear motions, resulting in a stronger impact on junction conductance. It's important to note that the slight difference in energy for the simulated vibrational mode when compared between **Fig. 4A** and **Fig. 4B** is due to the introduction of a top gold electrode in the simulation of the latter, as depicted in **fig. S9**. The energies, and spatial distribution of these vibrational modes closely agree with the features measured experimentally with STM-IETS. **Fig. 4, C-H** presents the simulated nuclear motions within CNAr$^{Mes2}$ in response to the molecular vibrational modes imaged in **Fig. 3, B-G**. The low-energy modes I-IV (**Fig. 4, C-F**) involve the frustrated rotational motion of different portions of CNAr$^{Mes2}$. It is noteworthy that we attribute mode IV to the superposition of 3 vibrations (**Fig. 4F**) of CNAr$^{Mes2}$ due to their similar energies. The hindered rotation of central aryl ring (**Fig. 4F, left**) and the stretching motion of the para-methyl groups (**Fig. 4F, middle and right**) together contribute to the bullseye feature in the d$^2$I/dV$^2$ mapping at 33.4 mV (**Fig. 3E**). The vibration at 48.5 meV is the bouncing motion (**Fig. 4G**) of CNAr$^{Mes2}$ on the surface and the 67.7 meV mode concerns the twisting of four ortho-methyl groups (**Fig. 4H**).

**Impact of Steric Effects on the Chemical Properties of CNAr$^{Mes2}$ on Au(111)**

The energies of the vibrational modes we observed vary in response to the molecular adsorption at different sites (**Fig. 5, A-C**), highlighting the effects of steric pressure and interference on molecular properties. At the step edge, the bouncing vibration (mode V) shows a blueshift compared to CNAr$^{Mes2}$ ligands adsorbed on the basal plane (**Fig. 5, D and E**), in good agreement with the theoretical simulation in **Fig. 4B**. This vibration involves the stretching motion of the carbon-metal bond. The increase in vibrational energy is consistent with a strengthened carbon-metal interaction at the step edge, resulting from the reduced out-of-plane steric repulsion between side arene groups and the metal substrate. In contrast, modes below 35 meV are largely absent in the spectra taken over the molecule at the step edge (**Fig. 5H**), as they involve frustrated rotation of the molecule in the polar direction and are likely quenched by the geometric limitation at the step edge. The simulated tunneling current captures the quenching of frustrated rotational modes I and III at the step edge, but the impact on modes II and IV is smaller compared to the experimental observation. We attribute this discrepancy to the selection rule of STM-IETS,(*65, 66*) which remains not fully understood.

A major difference between a molecule residing in a nano-environment and the ones in solution or a condensed phase arises from their intermolecular interactions. For m-terphenyl ligands, intermolecular interaction is also a critical determinant of steric pressure. These implications include the subtle energetic balance between attractive ligand-ligand interactions (i.e. due to van der Waals forces) and the central attractive forces that dictate ligand surface-binding. However, precise imaging and spectroscopic observation of these intermolecular interactions are extremely rare, despite their significant implications for understanding chemical binding to surfaces. Notably, we have observed distinct modifications in the vibrational features resulting from ligand-ligand interactions. For instance, the formation of a dimer between two molecules (as depicted in **Fig. 5C**) leads to an attractive interaction between the para-methyl group and the aryl ring of the adjacent molecule. Consequently, this interaction releases steric pressure on one side of the molecule while intensifying it on the other side, as shown in the theoretical simulation in **Fig. 4, I and J**. Experimentally, the redistribution of the steric pressure results in observed redshifts in the bouncing mode V of the CNAr$^{Mes2}$ ligands (**Fig. 5, D and E**), which can be attributed to a reduction in the ligand-surface bonding interaction as a consequence of strengthened ligand-ligand intermolecular interactions.(*67*) More strikingly, a clear splitting of the methyl twisting mode VI is observed (**Fig. 5, F and G**), resulting from the asymmetric steric pressure applied on the different ortho-methyl groups in the interacting ligands. Based on these results, it is likely that intermolecular interaction between ligands may diminish the strength of ligand-surface interactions, especially in low ligand-coverage scenarios.

**Conclusions**

In summary, we investigated the impact of steric interference on the surface adsorption behavior of m-terphenyl isocyanide ligands with STM. Additionally, we evaluated the effect of steric pressure on molecular vibrational properties through IETS and computational simulation. Our results unambiguously show that the steric repulsion applied on the individual $CNAr^{Mes2}$ ligands are reduced when adsorbed on a convex surface, leading to a site-selective molecular binding. The sub-molecular scale characterization provides detailed insights into the unique chemical environment experienced by each individual ligand, revealing information that was previously inaccessible through ensemble measurements. Specifically, the vibrational characterization of (i) individual $CNAr^{Mes2}$ ligands adsorbed on surface sites with varying curvatures and (ii) a ligand dimer reveals the influence of ligand-surface and intermolecular interactions on modifying the steric response of the constituent molecules. This information holds particular significance in the field of nanoscience, where even minor variations in the nanoscale chemical environment have been demonstrated to significantly impact molecular behavior.

Overall, the unprecedented molecular-scale study of binding dynamics of ligand using the steric encumbrance as a design principle on Au(111) brings us one step closer to decipher that in real and nonideal ligand-surface systems. Moreover, the factors demonstrated to strongly influence the binding behavior of $CNAr^{Mes2}$ have the potential to facilitate the rational design of other ligands targeting diverse surface topologies.

**Acknowledgement:**

**Funding:** The authors acknowledge the use of facilities and instrumentation supported by National Science Foundation (NSF) through the UC San Diego Materials Research Science and Engineering Center (UCSD MRSEC) with Grant No. DMR-2011924. This material is based upon work supported by the NSF under Grant CHE-2303936 and No. DMR-2011924. This work used the Expanse supercomputer at the San Diego Supercomputing Center through allocation DDP381 from the Advanced Cyberinfrastructure Coordination Ecosystem: Services & Support (ACCESS) program, which is supported by NSF grants No. 2138259, No. 2138286, No. 2138307, No. 2137603, and No. 2138296.

**Author Contributions:**

Conceptualization, funding acquisition and supervision: ART, TAP, JSF, SL

Ligand Synthesis: KPB, JSF

STM Measurement: LB, QZ, SL

Computational Simulation: SJ, AC, AD, ZW, YZ, TAP

All authors contributed to the writing and revision of the manuscript.

**Competing Interests:** The authors declare no competing interests.

**Data and materials availability:** All data are available in the main text or the supplementary materials.

**Supplementary Materials**

Materials and Methods

Supplementary Text

Figs. S1 to S9

Table S1

Movies S1 to S3

References (23,40,57,68-85)

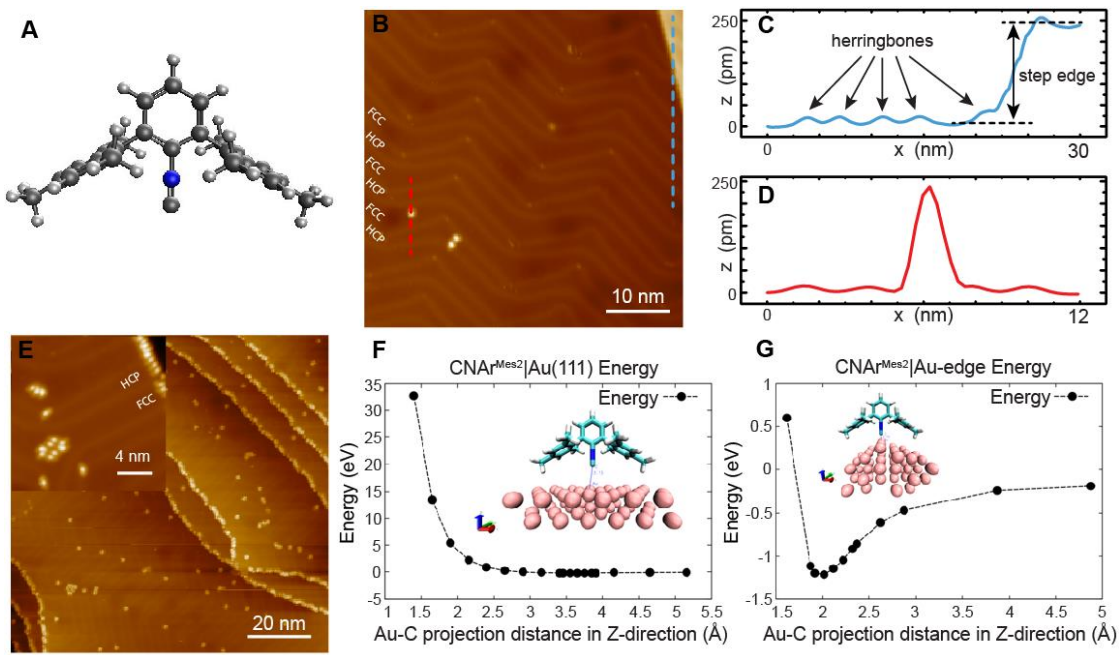

**Fig. 1. Thermally induced diffusion of CNAr$^{Mes2}$ ligand on Au(111).** (**A**) Ball-and-stick model of CNAr$^{Mes2}$, with N, C, and H atoms shown in blue, dark grey, and light grey, respectively. (**B**) Topographic image of randomly adsorbed CNAr$^{Mes2}$ (an isolated one and a dimer) on Au(111) at 5K. The FCC and HCP domains are labeled for clarity. (**C**) Line profile across the blue dashed line in (B), showing the diverse landscapes on Au(111). (**D**) Line profile across the isolated CNAr$^{Mes2}$ in (B). (**E**) Topographic image of the surface after annealing to room temperature, with zoom-in inset showing CNAr$^{Mes2}$ in clusters, on herringbone elbow sites, and at the step edge. (**F** and **G**) Simulated adsorption energy landscapes of CNAr$^{Mes2}$ on Au(111) at the planar (F) and curved surface (G). Imaging parameters were set to -1 V, 100 pA for (B), 1 V, 100 pA for (E) and -500 mV, 50 pA for inset of (E).

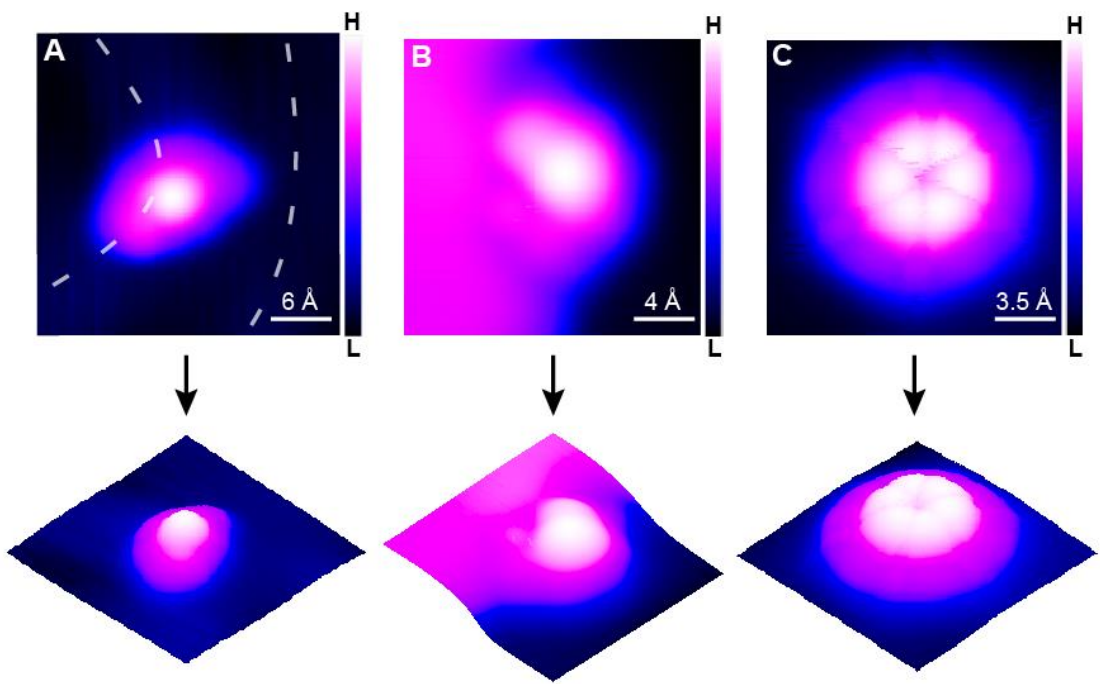

**Fig. 2. Visualizing the impact of steric effects on the adsorption structure of single CNAr$^{Mes2}$.** (**A**)-(**C**) Topographic images of individual CNAr$^{Mes2}$ on the herringbone elbow site (A), step edge (B), and planar surface (C). The white dashed curves in (A) indicate the herringbones. Imaging parameters were set to (A) -50 mV, 20 pA, and (B and C) -85 mV, 100 pA.

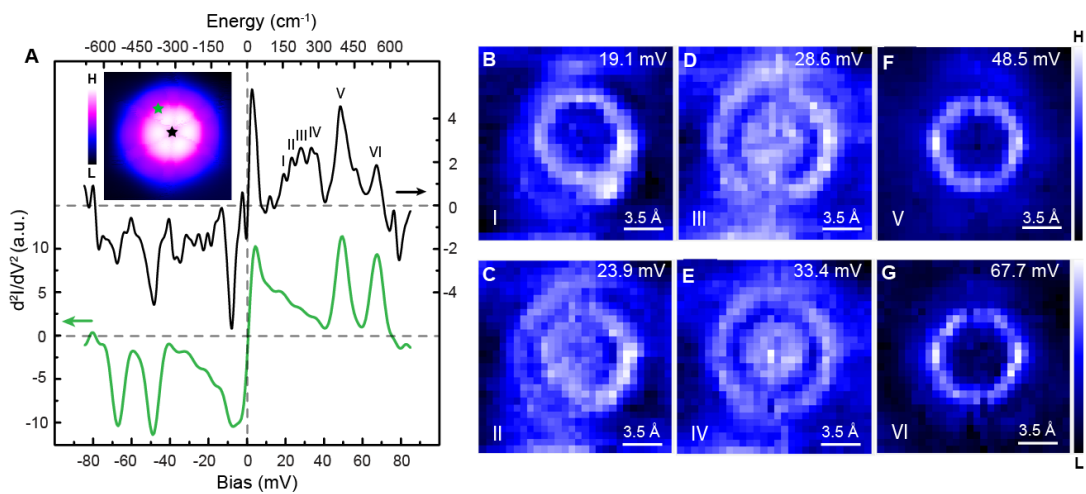

**Fig. 3. Vibrational characterization of individual CNAr$^{Mes2}$.** (**A**) IETS spectra of the molecule (inset) at two different positions indicated by black and green stars, respectively. The spectra are vertically shifted for clarity. Imaging and spectroscopy parameters were set to -85 mV, 100 pA and -85 mV, 1.2 nA, respectively. (**B**)-(**G**) d$^2$I/dV$^2$ mapping of the same CNAr$^{Mes2}$ at six different bias voltages, ranging from 19.1 mV to 67.7 mV. Imaging parameters for the mapping were set to -67.7 mV, 400 pA.

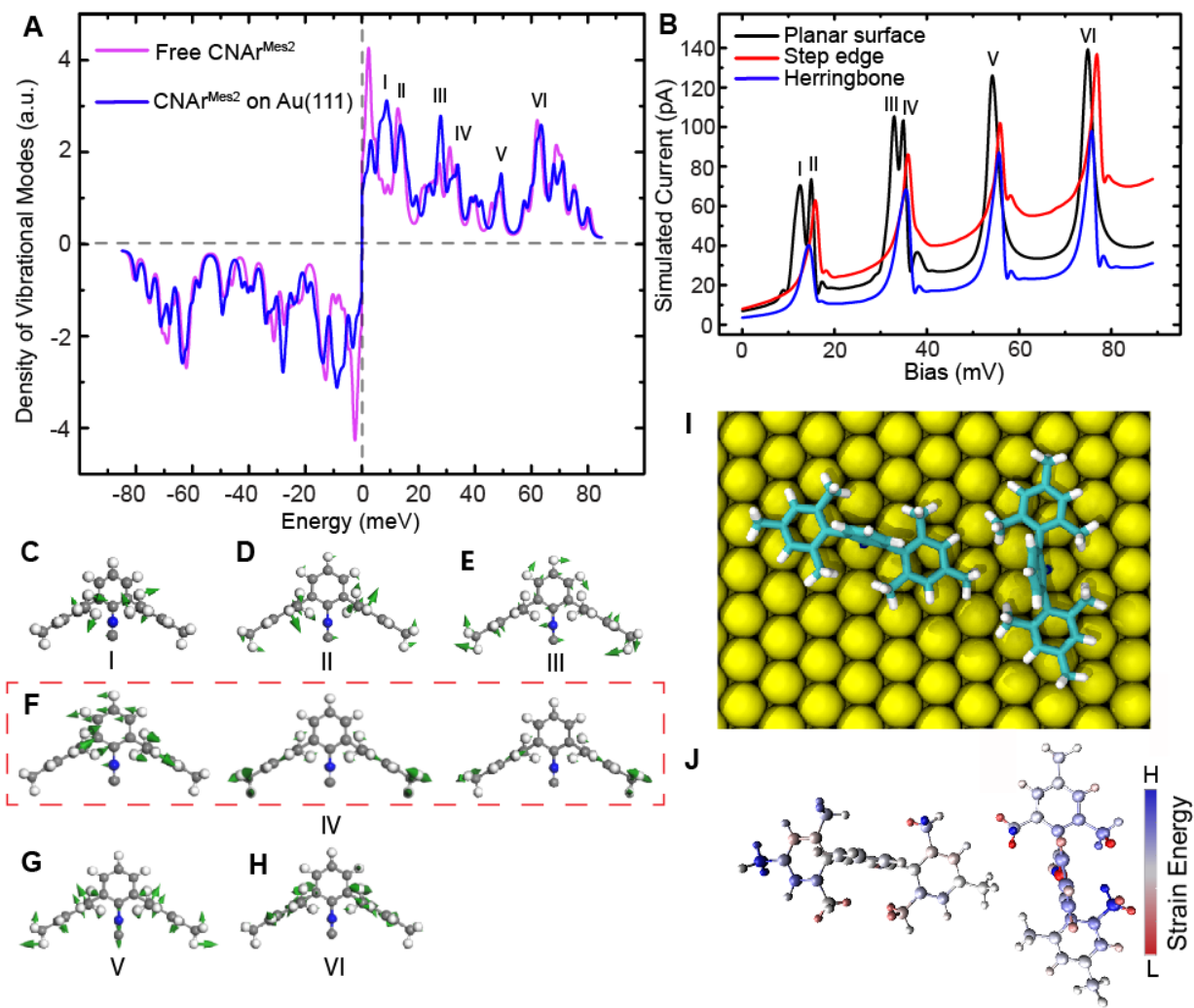

**Fig. 4. Theoretically calculated vibrational modes of individual CNAr^Mes2 and strain distribution within a CNAr^Mes2 dimer.** (**A**) Calculated density of vibrational modes of free CNAr$^{Mes2}$ (magenta) and CNAr$^{Mes2}$ on Au(111) (blue). (**B**) Calculated IV curve of CNAr$^{Mes2}$ on planar surface (black), at the step edge (red) and on herringbone (blue) of Au(111). (**C**)-(**H**) Simulated nuclear motions within a free CNAr$^{Mes2}$. (**I**) Top view of a simulated CNAr$^{Mes2}$ dimer on Au(111). (**J**) Top view of the simulated strain energy distribution within the CNAr$^{Mes2}$ dimer in (I). Blue (red) means increased (decreased) strain compared to the monomer.

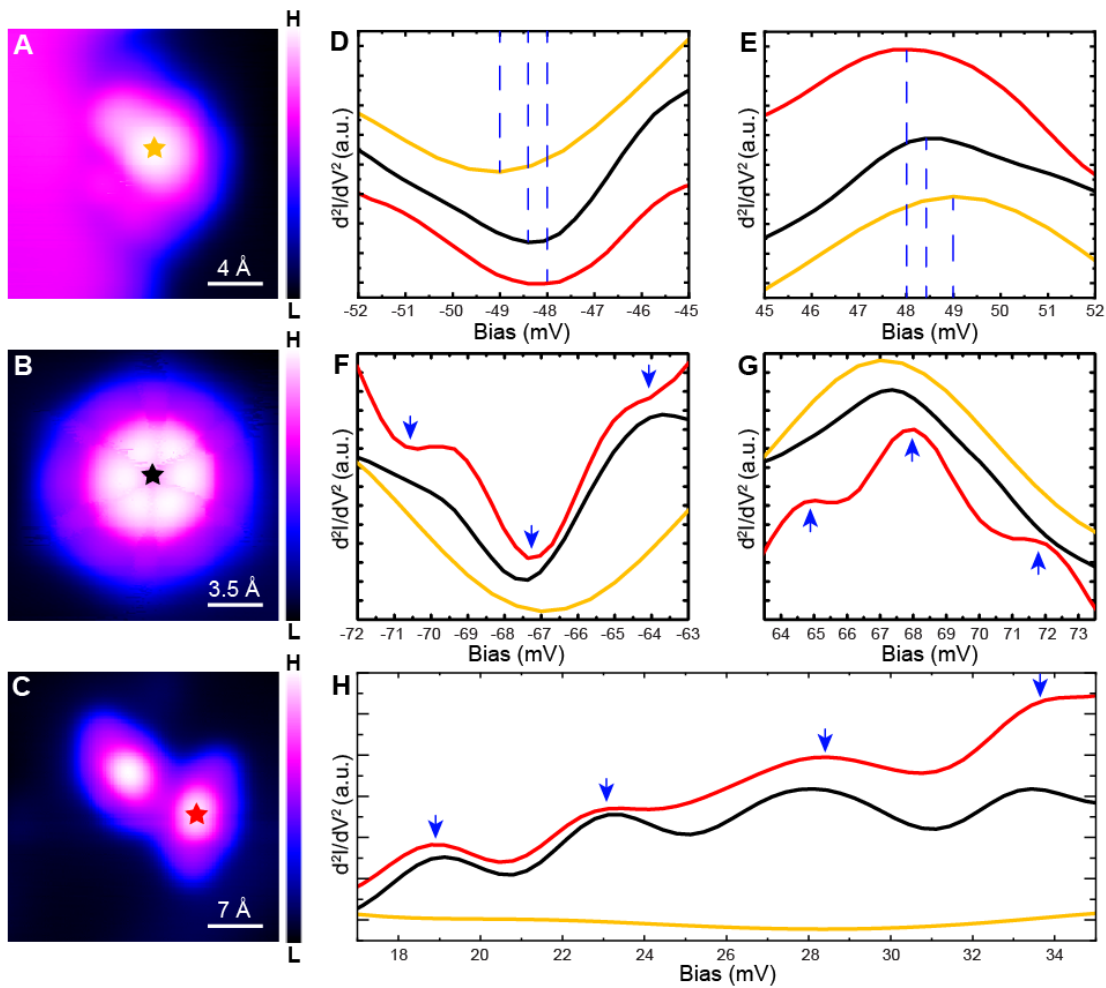

**Fig. 5. Effect of steric pressure and intermolecular interaction on molecular vibrations of CNAr$^{Mes2}$.** (**A**)-(**C**) Topographic images of CNAr$^{Mes2}$ at the step edge (A), on the planar surface (B), and in a dimer configuration (C). (**D**)-(**H**) Comparison of IETS spectra demonstrating vibrational shifts (D and E), vibrational splitting (F and G), and vibrational quenching (H) of CNAr$^{Mes2}$ ligands. The positions where the IETS spectra in (D)-(H) were collected are marked with orange, black, and red stars in (A)-(C). Imaging and spectroscopy parameters were set to -85 mV, 100 pA for (A) and (B), -500 mV, 30 pA for (C), and -85 mV, 1.2 nA for (D)-(H).